\documentclass[12pt]{article}
\usepackage{amsmath}
\usepackage{amsfonts}   
\usepackage{amssymb}
\begin{document}
\date{}
\title{\textbf{Lie algebraic Noncommutative Gravity}}
\author{{Rabin Banerjee}$^{a}$\thanks{E-mail: rabin@bose.res.in}, \ {Pradip Mukherjee}$^{b,c}$\thanks{E-mail: pradip@bose.res.in} \ and {Saurav Samanta}$^{a,}$\thanks{E-mail: saurav@bose.res.in}\\
$^{a}$ {\textit{S.~N.~Bose National Centre for Basic Sciences,}}\\{\textit{JD Block, Sector III, Salt Lake, Kolkata-700098, India}}\\[0.3cm]
$^{b}$ {\textit{Presidency College}}\\{\textit{86/1 College Street, Kolkata-700073, West-Bengal, India}}\\[0.3cm]
$^{c}$ {Also Visiting Associate,}\\
{\textit{ S.~N.~Bose National Centre for Basic Sciences,}}\\
{\textit{ JD Block, Sector III, Salt Lake, Kolkata-700098, India}}\\
{\textit{ and IUCAA,}}
\\{\textit{Post Bag 4, Pune University Campus,}}
\\{\textit{ Ganeshkhind, Pune 411007, India}}\\[0.3cm]
}
\maketitle
                                                                                
\begin{quotation}
\noindent \normalsize 
We exploit the Seiberg -- Witten map technique to formulate the theory of gravity defined on a Lie algebraic noncommutative space time. Detailed expressions of the Seiberg -- Witten maps for the gauge parameters, gauge potentials and the field strengths have been worked out. Our results demonstrate that notwithstanding the introduction of more general noncommutative structure there is no first order
correction, exactly as happens for a canonical (i.e. constant) noncommutativity.
\end{quotation}
\section{Introduction}
There is a broad consensus that general relativity and ordinary differential geometry should be replaced by noncommutative (NC) geometry at some point between currently accessible energies of about 1 -- 10 Tev (~projected for the LHC at CERN ) and the Planck scale, which is $10^{15}$ times higher. Indeed, formulation of the general theory of relativity in the NC perspective is considered to be a necessity for quantizing gravity\cite{szabo1}. The principal obstacle in this formulation comes from negotiating general coordinate invariance. Different approaches to the problem can be broadly classified on the manner in which the diffeomorphism invariance of general relativity has been treated in the NC setting. In \cite{chams} a deformation of Einstein's gravity was studied using a construction based on gauging the noncommutative $SO(4,1)$ de sitter group and the Seiberg -- Witten (SW) map \cite{SW} with subsequent contraction to $ISO(3,1)$. Construction of a noncommutative gravitational theory was proposed based on a twisted diffeomorphism algebra \cite{Aschierie,kur}. On the other hand it has been shown that the theory can be formulated basing on true physical symmetries \cite{CK1} by resorting to a class of restricted coordinate transformations that preserve the NC algebra. The restriction corresponds to the formulation of NC gravity in the context of unimodular gravity \cite{Ein}. This is sometimes referred as the minimal formulation of NC gravity. All these observations are valid only for a constant noncommutative parameter $\theta$. There is also an analogous discussion for a covariantly constant $\theta$ \cite{hari}.

      A remarkable feature is that there is no first order correction term in $\theta$ for the various theories of noncommutative gravity for constant $\theta$\cite{chams,Aschierie,MS}. It was also conjectured \cite{MS,Deli} that the underlying symmetry of the commutative space -- time is instrumental in the vanishing of the first order correction. Nontrivial contribution starts from the second order term \cite{chams,Aschierie,CK2}. However, considering the various estimates of the size of the noncommutative parameter \cite{Ani}, the noncommutative correction to gravity appears to be insignificant, at least in the intermediate energy regime below the Planck scale.

 All the formulations of noncommutative gravity are usually discussed assuming the canonical noncommutative algebra 
\begin{eqnarray}
[\hat{x}^{\mu},\hat{x}^{\nu}]=i\theta^{\mu\nu}
\label{constant algebra}
\end{eqnarray}
where $\theta^{\mu\nu}$ is a constant antisymmetric two index object. The question that naturally appears is whether the vanishing of the order $\theta$ correction is due to this restriction. Perhaps a more general noncommutative structure might lead to order $\theta$ effects.

The motivation of the present paper is to address this issue of $\mathcal{O}(\theta)$ effects for a more general type of noncommutativity. The obvious generalization beyond a constant $\theta$ is to take a noncommutative parameter which is linear in the coordinates. This naturally leads to a noncommutativity of the form
\begin{eqnarray}
[\hat{x}^{\mu},\hat{x}^{\nu}]&=&i\theta^{\mu\nu}(\hat{x})\nonumber\\
&=&i\theta f^{\mu\nu}_{ \ \ \ \lambda}\hat{x}^{\lambda}
\label{lie}
\end{eqnarray}
where $f^{\mu\nu}_{ \ \ \ \lambda}$ are the structure constants. Later we will find that for consistency within our approach these constants assume a Lie algebraic structure so that $f_{\mu\nu\lambda}$ is antisymmetric in all the three indices. We prefer to carry out our analysis in the framework of the minimal theory \cite{CK1} where the true symmetries are manifest.
This has a distinct advantage in that it uses the tetrad formalism where the general coordinate invariance is viewed as a local symmetry implemented by the tetrad as the gauge field along with the local Lorentz invariance $(SO(3,1))$ generated by the spin connection fields. This enables one to use the machinery of noncommutative gauge theories elaborately developed in the literature \cite{SW,W1,W2,W3}. The celebrated Seiberg -- Witten map technique \cite{SW} can be used to cast the theory of noncommutative gravity as a perturbative theory in the noncommutative parameter $\theta$. Such maps have been exhaustively available for the canonical structure \cite{W3}. Constructions valid for general coordinate dependent noncommutativity have been given \cite{Behr} but these are not directly suitable for our calculation. We will thus require to develop the appropriate maps for the Lie -- algebra valued coordinate dependent non -- commutative structure in the sequel. Thus apart from the principal motivation of looking for a first order effect our study of noncommutative gravity will be intrinsically interesting for the issues of constructing these maps and the noncommutative gauge -- invariances involved therein.   

     Before proceeding further it is useful to dwell on the organisation of the paper. In the next section we will discuss about the class of the general coordinate transformations which is consistent with the general noncommutative algebra (\ref{lie}). In section -- 3 the requirements of the noncommutative gauge invariances will be analysed. Appropriate SW maps will be constructed for the gauge potentials and field strength tensors valid for the Lie algebraic noncommutativity. In section -- 4 the main results of noncommutative gravity in the tetrad formalism will be  presented. We will show that there is no first order correction in the commutative equivalent action of NC gravity. We will conclude in section --5.
 
\section{General coordinate transformation for noncanonical noncommutative space}
The formulation of gravity on NC space time poses problems. This is seen by considering the general coordinate transformation,        
\begin{eqnarray}
\hat{x}^{\mu}\rightarrow \hat{x}'^{\mu}=\hat{x}^{\mu}+\hat{\xi}^{\mu}(\hat{x})
\label{xxi}
\end{eqnarray}
and realising that, for arbitrary $\hat{\xi}^{\mu}(\hat{x})$, it is not compatible with the algebra (\ref{lie}). However, as in the canonical case\cite{CK1}, it is possible to find a restricted class of coordinate transformations (\ref{xxi}) which preserves the Lie -- algebraic noncommutative algebra. To demonstrate this assertion in a convenient way and also for further developments it will be appropriate to exploit the Weyl -- correspondence \cite{Weyl} to work in the deformed phase space with the ordinary multiplication substituted by the corresponding star product. A short review of the formalism is thus appropriate at this stage.
	 
	 The noncommutative coordinates $\hat{x}^{\mu}$ satisfying (\ref{lie}) are the generators of an associative algebra ${\cal{A}}_x$. According to the Weyl correspondence we can associate an element of ${\cal{A}}_x$ with a function $f(x)$ of classical variables $x^{\mu}$ by the unique prescription
\begin{equation}
W(f) = \frac{1}{(2\pi)^2} \int d^4 k e^{ik_{\mu}{\hat {x}}^{\mu}}{\tilde{f}}(k)
\label{Weyl}
\end{equation}
where ${\tilde{f}}(k)$ is the Fourier transform of $f(x)$. The *-product between two classical functions $f(x)$ and $g(x)$ is denoted by $f*g$ and is defined by the requirement
\begin{equation}
W(f)W(g) = W(f*g).
\end{equation}
 The elements of ${\cal{A}}_x$ can then be represented by ordinary functions with their product defined by the star product. When the generators satisfy the Lie structure the star product is explicitly given by \cite{W1}
 \begin{eqnarray}
f(x)*g(x)=e^{\frac{i}{2}x^{\lambda}g_{\lambda}(i\frac{\partial}{\partial x'},i\frac{\partial}{\partial x''})}f(x')g(x'')|_{(x',x'')\rightarrow x}
\label{star}
\end{eqnarray}
where $g_{\lambda}$ is defined by,
\begin{eqnarray}
e^{ik_{\lambda}\hat{x}^{\lambda}}e^{ip_{\lambda}\hat{x}^{\lambda}}=e^{i\{k_{\lambda}+p_{\lambda}+\frac{1}{2}g_{\lambda}(k,p)\}\hat{x}^{\lambda}}
\end{eqnarray}
Using the Baker-Campbell-Hausdorff formula,
\begin{equation}
e^Ae^B = e^{A+B+\frac{1}{2}[A,B]+\frac{1}{12}([A,[A,B]]-[B,[A,B]])-\frac{1}{48}([B,[A,[A,B]]]+[A,[B,[A,B]]])...}
\end{equation}
 an explicit form of $g_{\lambda}(k,p)$ is obtained
\begin{eqnarray}
g_{\lambda}(k,p)&=&-\theta k_{\mu}p_{\nu}f^{\mu\nu}_{ \ \ \ \lambda}+\frac{1}{6}\theta^2k_{\mu}p_{\nu}(p_{\sigma}-k_{\sigma})f^{\mu\nu}_{ \ \ \ \delta}f^{\delta\sigma}_{ \ \ \ \lambda}\nonumber\\&&+\frac{1}{24}\theta^3(p_{\sigma}k_{\beta}+k_{\sigma}p_{\beta})k_{\mu}p_{\nu}f^{\mu\nu}_{ \ \ \ \delta}f^{\delta\sigma}_{ \ \ \ \alpha}f^{\alpha\beta}_{ \ \ \ \lambda}+...
\end{eqnarray}

        After this brief digression we now turn to the determination of the restriction on the transformations (\ref{xxi}) such that the noncommutative algebra (\ref{lie}) is preserved. From (\ref{xxi}), using the Weyl correspondence, we get
\begin{eqnarray}
[x'^{\mu},x'^{\nu}]_*=[x^{\mu},x^{\nu}]_*+[x^{\mu},\hat{\xi}^{\nu}(x)]_*+[\hat{\xi}^{\mu}(x),x^{\nu}]_*+\mathcal{O}(\hat{\xi}^2).
\end{eqnarray}

Using the formula (\ref{star}) one derives the following relation\cite{W1},
\begin{eqnarray}
[x^{\mu},f(x)]_*=i\theta f^{\mu\nu}_{ \ \ \ \lambda}x^{\lambda}\frac{\partial f}{\partial x^{\nu}}.
\label{formula}
\end{eqnarray}
It is then straightforward to find, using (\ref{formula}), that in order to preserve (\ref{lie}), $\xi^{\mu}$ must satisfy the condition,
\begin{eqnarray}
i\theta f^{\mu\sigma}_{ \ \ \ \lambda}x^{\lambda}\frac{\partial \hat{\xi}^{\nu}}{\partial x^{\sigma}}-i\theta f^{\nu\sigma}_{ \ \ \ \lambda}x^{\lambda}\frac{\partial \hat{\xi}^{\mu}}{\partial x^{\sigma}}=i\theta f^{\mu\nu}_{ \ \ \ \lambda}\hat{\xi}^{\lambda}(x).
\end{eqnarray}
A nontrivial solution of the above equation is given by,
\begin{eqnarray}
\hat{\xi}^{\mu}(x)=f^{\mu\alpha}_{ \ \ \ \beta}x^{\beta}\partial_{\alpha}g(x).
\label{restrtrans}
\end{eqnarray}
This can be checked by using the Jacobi identity following from (\ref{lie})
\begin{eqnarray}
f^{\mu\nu}_{ \ \ \ \sigma}f^{\sigma\lambda}_{ \ \ \ \delta}+f^{\nu\lambda}_{ \ \ \ \sigma}f^{\sigma\mu}_{ \ \ \ \delta}+f^{\lambda\mu}_{ \ \ \ \sigma}f^{\sigma\nu}_{ \ \ \ \delta} = 0.
\label{jacobi-lie}
\end{eqnarray}
Equation (\ref{restrtrans}) gives the restricted class of general coordinate transformations under which the noncommutative algebra (\ref{lie}) is preserved. From (\ref{restrtrans}) we find that 
\begin{eqnarray}
\partial_{\mu}\hat{\xi}^{\mu}(x) = 0.\nonumber
\end{eqnarray}
The Jacobian of the transformations (\ref{xxi}) is then unity. In other words the transformations are volume preserving. The corresponding theory thus belongs to the noncommutative version of unimodular gravity.

\section{Seiberg-Witten map for Lie algebraic noncommutativity}
In the framework of tetrad gravity two gauge symmetries need to be implemented -- one is the translation of the tetrad while the other is the local homogeneous Lorentz transformations. We thus consider a noncommutative gauge theory, valued in $ISO(3,1)$ Lie algebra. The Seiberg -- Witten (SW) maps for the non-Abelian noncommutative gauge fields where the noncommutative coordinates satisfy the canonical algebra are elaborately worked out in the literature \cite{SW,W1,W2,W3,RK}. But the corresponding results for Lie algebraic noncommutative structure are only sketched \cite{W1,W2,Behr}. A comprehensive analysis will thus be appropriate at this stage which will also be useful for the subsequent calculations. Also all our SW maps are valid upto first order in $\theta$ since we are interested only in the first order effects.

  A non-Abelian gauge theory in the noncommutative space carries two algebraic structures, the associative algebra $\mathcal{A}_x$ whose generators are the noncommutative coordinates
$\hat{x}^{\mu}$ and the nonabelian Lie algebra $\mathcal{A}_T$, the hermitian generators $T^{a}$ of which
satisfy the algebra
\begin{eqnarray}
[T^a,T^b]=il^{ab}_{ \ \ \ c}T^c.
\end{eqnarray}

In classical coordinate space where $\mathcal{A}_x$ is commutative we can define the field $\psi(x)$ in some representation of the Lie group and introduce the gauge potential $A_i(x)=A_{ia}(x)T^a$ in the usual way to make the symmetry local. The appropriate gauge transformations are
\begin{eqnarray}
\delta_{\alpha}\psi(x)=i\alpha(x)\psi(x), \  \  \  \ \alpha(x)=\alpha_a(x)T^a
\end{eqnarray}
and
\begin{eqnarray}
\delta_{\alpha}A_{\mu}(x)=\partial_{\mu}\alpha(x)+i[\alpha(x),A_{\mu}(x)].
\end{eqnarray} 
Note that the commutator of two gauge transformations is another transformation in the same gauge group
\begin{eqnarray}
(\delta_{\alpha}\delta_{\beta}-\delta_{\beta}\delta_{\alpha})\psi(x)=\delta_{-i[\alpha,\beta]}\psi(x)
\label{closure}
\end{eqnarray}

In the noncommutative space, on the other hand, the closure (\ref{closure}) does not hold \cite{W1} within the Lie algebra but is satisfied in the enveloping algebra. Thus the noncommutative field $\hat{\psi}(x)$ transforms as\cite{W2}
\begin{eqnarray}
\delta_{\hat{\alpha}}\hat{\psi}(\hat{x})=i\hat{\alpha}(\hat{x})\hat{\psi}(\hat{x})
\end{eqnarray}
which is written in * product formalism as
\begin{eqnarray}
\delta_{\hat{\alpha}}\hat{\psi}(x)=i\hat{\alpha}(x)*\hat{\psi}(x)
\label{delsi}
\end{eqnarray}
where the gauge parameter $\hat{\alpha}(x)$ is in the enveloping algebra \cite{W2}
\begin{eqnarray}
\hat{\alpha}(x)=\hat{\alpha}_a(x):T^a:+\hat{\alpha}^1_{ab}(x):T^aT^b:+...+\hat{\alpha}^{n-1}_{a_1...a_n}(x):T^{a_1}...T^{a_n}:+...
\end{eqnarray}
where
\begin{eqnarray}
:T^a:&=&T^a\\
:T^aT^b:&=&\{T^a,T^b\}\\
:T^{a_1}...T^{a_n}:&=&\frac{1}{n!}\sum_{\pi\in S_n}T^{a_{\pi(1)}}...T^{a_{\pi(n)}}
\end{eqnarray}
All these infinitely many parameters $\hat{\alpha}^{n-1}_{a_1...a_n}(x)$ depend only on the commutative gauge parameter $\alpha(x)$, the gauge potential $A_{\mu}(x)$ and on their derivatives. We denote this as $\hat{\alpha}\equiv\hat{\alpha}(\alpha(x),A(x))$. Then it follows from (\ref{delsi}) that the variation of $\hat{\psi}$ is expressed as{\footnote{We are following the notation of \cite{W3} in the sense that $\delta_{\hat{\alpha}}\hat{\psi}$ is now written as $\delta_{\alpha}\hat{\psi}$ since $\hat{\alpha}$ is expressed as a function of $\alpha$.}},
\begin{eqnarray}
\delta_{\alpha}\hat{\psi}(x)=i\hat{\alpha}(\alpha(x),A(x))*\hat{\psi}(x).
\end{eqnarray}
Now we impose the requirement of closure,
\begin{eqnarray}
(\delta_{\alpha}\delta_{\beta}-\delta_{\beta}\delta_{\alpha})\hat{\psi}(x)=\delta_{-i[\alpha,\beta]}\hat{\psi}(x).
\end{eqnarray}
Then the above equation is written in the explicit form
\begin{eqnarray}
i\delta_{\alpha}\hat{\beta}(\beta,A)-i\delta_{\beta}\hat{\alpha}(\alpha,A)+\hat{\alpha}(\alpha,A)*\hat{\beta}(\beta,A)-\hat{\beta}(\beta,A)*\hat{\alpha}(\alpha,A)\nonumber\\
=i(\widehat{-i[\alpha,\beta]})(-i[\alpha,\beta],A)
\end{eqnarray} 
Expanding in $\theta$ we write,
\begin{eqnarray}
\hat{\alpha}(\alpha,A)=\alpha+\theta\alpha^1(\alpha,A)+\mathcal{O}(\theta^2)
\end{eqnarray}
To first order we obtain,
\begin{eqnarray}
i\delta_{\alpha}\beta^1(\beta,A)-i\delta_{\beta}\alpha^1(\alpha,A)+[\alpha,\beta^1(\beta,A)]-[\beta,\alpha^1(\alpha,A)]\nonumber\\
-i(-i[\alpha,\beta])^1(-i[\alpha,\beta],A)=-\frac{i}{2}f^{\mu\nu}_{ \  \ \lambda}x^{\lambda}\{\partial_{\mu}\alpha,\partial_{\nu}\beta\}
\end{eqnarray}
The solution is given by
\begin{eqnarray}
\theta\alpha^1(\alpha,A)=\frac{1}{4}\theta^{\mu\nu}\{\partial_{\mu}\alpha,A_{\nu}\}
\label{29}
\end{eqnarray}
where $\theta^{\mu\nu}$ is defined in (\ref{lie})

Similarly we can expand the field $\hat{\psi}$ also as,
\begin{eqnarray}
\hat{\psi}(A)=\psi+\theta\psi^1(A)+\mathcal{O}(\theta^2).
\end{eqnarray}
To first order,
\begin{eqnarray}
\delta_{\alpha}\psi^1(A)
=i\alpha\psi^1(A)+i\alpha^1(\alpha,A)\psi-\frac{1}{2}f^{\mu\nu}_{ \  \ \lambda}x^{\lambda}\partial_{\mu}\alpha\partial_{\nu}\psi.
\end{eqnarray}
Its solution is
\begin{eqnarray}
\theta\psi^1(A)=-\frac{1}{2}\theta^{\mu\nu}A_{\mu}\partial_{\nu}\psi+\frac{i}{4}\theta^{\mu\nu}A_{\mu}A_{\nu}\psi.
\end{eqnarray}
The noncommutative gauge potential $\hat{A}_{\mu}$ is most naturally introduced by
the covariant coordinate $\hat{X}^{\mu}$ approach developed by Wess and collaborators \cite{W1}. $\hat{X}^{\mu}$ is defined in the following way
\begin{eqnarray}
\hat{X}^{\mu}(\hat{x})=\hat{x}^{\mu}+\hat{A}^{\mu}(\hat{x})
\end{eqnarray}
which when acts on $\hat{\psi}$ transforms covariantly\cite{W2}, i.e.
\begin{eqnarray}
\delta_{\alpha}\hat{X}^{\mu}(\hat{x})\hat{\psi}(\hat{x})=i\hat{\alpha}(\hat{x})\hat{X}^{\mu}(\hat{x})\hat{\psi}(\hat{x}).
\end{eqnarray}
This requirement gives the transformation of $\hat{A}^{\mu}$
\begin{eqnarray}
\delta_{\alpha}\hat{A}^{\mu}(\hat{x})=-i[\hat{x}^{\mu},\hat{\alpha}(\hat{x})]+i[\hat{\alpha}(\hat{x}),\hat{A}^{\mu}(\hat{x})].
\end{eqnarray}
In the star product formalism
\begin{eqnarray}
\delta_{\alpha}\hat{A}^{\mu}(x)=-i[x^{\mu},\hat{\alpha}(x)]_*+i[\hat{\alpha}(x),\hat{A}^{\mu}(x)]_*
=\theta^{\mu\rho}\frac{\partial}{\partial x^{\rho}}\hat{\alpha}+i[\hat{\alpha}(x),\hat{A}^{\mu}(x)]_*
\end{eqnarray}
The gauge potential $\hat{A}_{\mu}$ is defined through $\hat{A}^{\mu}$ as in the case of canonical noncommutativity \cite{W3}
\begin{eqnarray}
\hat{A}^{\mu}=\theta^{\mu\rho}\hat{A}_{\rho}.
\end{eqnarray}
Due to the coordinate dependence of the noncommutative structure $\theta^{\mu\rho}$ it is not possible to find the transformation of $\hat{A}_{\mu}$ in closed form but one can obtain results correct upto the required order in the noncommutative parameter.
To first order in $\theta$
\begin{eqnarray}
\delta_{\alpha}\hat{A}_{\mu}=\partial_{\mu}\hat{\alpha}+i[\hat{\alpha},\hat{A}_{\mu}]-\frac{1}{2}\theta^{\lambda\sigma}\{\partial_{\lambda}\hat{\alpha},\partial_{\sigma}\hat{A}_{\mu}\}-\frac{1}{2}\theta_{\mu\alpha}\theta^{\lambda\sigma}\partial_{\sigma}\theta^{\alpha\beta}\{\partial_{\lambda}\hat{\alpha},\hat{A}_{\beta}\}
\label{deltaA}
\end{eqnarray}
where $\theta_{\mu\alpha}$ is the inverse of $\theta^{\mu\alpha}$.

  To get the Seiberg -- Witten map for the gauge potential
we expand it in a perturbative series in      $\theta$
\begin{eqnarray}
\hat{A}_{\mu}(A)=A_{\mu}+\theta A^1_{\mu}(A)+\mathcal{O}(\theta^2)
\label{SWA}
\end{eqnarray}
Computing the gauge transformation of $\hat{A}_{\mu}$ from
the above using the corresponding transformation of the commutative potential and comparing with (\ref{deltaA})
we get
\begin{eqnarray}
\delta_{\alpha} A^1_{\mu}(A)=\partial_{\mu}\alpha^1(\alpha,A)+i[\alpha^1(\alpha,A),A_{\mu}]+i[\alpha,A^1_{\mu}(A)]-\frac{1}{2}f^{\nu\lambda}_{ \  \ \delta}x^{\delta}\{\partial_{\nu}\alpha,\partial_{\lambda}A_{\mu}\}
\end{eqnarray}
The solution to the last equation is
\begin{eqnarray}
\theta A^1_{\mu}(A)=-\frac{1}{4}\theta^{\nu\lambda}\{A_{\nu},\partial_{\lambda}A_{\mu}+F_{\lambda\mu}\}-\frac{1}{4}\theta_{\mu\nu}\theta^{\lambda\sigma}\partial_{\sigma}\theta^{\nu\delta}\{A_{\lambda},A_{\delta}\}
\label{seibergA}
\end{eqnarray}
where,
\begin{eqnarray}
F_{\mu\nu}=\partial_{\mu}A_{\nu}-\partial_{\nu}A_{\mu}-i[A_{\mu},A_{\nu}].
\end{eqnarray}
We then get the Seiberg -- Witten map for the gauge potential correct upto first order in $\theta$ from (\ref{SWA}) and (\ref{seibergA}) as
\begin{equation}
\hat{A}_{\mu}(A)= A_{\mu} - \frac{1}{4}\theta^{\nu\lambda}\{A_{\nu},\partial_{\lambda}A_{\mu}+F_{\lambda\mu}\}-\frac{1}{4}\theta_{\mu\nu}\theta^{\lambda\sigma}\partial_{\sigma}\theta^{\nu\delta}\{A_{\lambda},A_{\delta}\}.
\label{A1}
\end{equation}
Note that the last term on the r.h.s. is nonvanishing when $\theta^{\mu\nu}$ is coordinate dependent. In the limit of constant $\theta$ the usual SW map is retrieved\cite{SW}.

   Our next task is to construct the Seiberg -- Witten maps
for the Yang -- Mills field $\hat{F}_{\mu\nu}$.
We first define a second rank tensor
\begin{eqnarray}
\hat{F}^{\mu\nu}(\hat{x})&=&-i\left([\hat{X}^{\mu}(\hat{x}),\hat{X}^{\nu}(\hat{x})]-i\theta f^{\mu\nu}_{ \ \ \ \lambda}\hat{X}^{\lambda}(\hat{x})\right).\nonumber
\end{eqnarray}
This is written in the * product notation as
\begin{eqnarray}
\hat{F}^{\mu\nu}(x)=-i\left([x^{\mu},\hat{A}^{\nu}(x)]_*-[x^{\nu},\hat{A}^{\mu}(x)]_*+[\hat{A}^{\mu}(x),\hat{A}^{\nu}(x)]_*-i\theta f^{\mu\nu}_{ \ \ \ \lambda}\hat{A}^{\lambda}(x)\right)
\label{f}
\end{eqnarray}
which transforms covariantly as
\begin{eqnarray}
\delta_{\alpha}\hat{F}^{\mu\nu}=i[\hat{\alpha},\hat{F}^{\mu\nu}]_*
\label{trans}
\end{eqnarray}
 This second rank tensor allows us to define the Yang -- Mills $\hat{F}_{\mu\nu}$ through
\begin{eqnarray}
\hat{F}^{\mu\nu}=\theta^{\mu\lambda}\theta^{\nu\sigma}\hat{F}_{\lambda\sigma}
\label{45}
\end{eqnarray}
From eq. (\ref{f}), we get the following expression for $\hat{F}_{\mu\nu}$, 
\begin{eqnarray}
\hat{F}_{\mu\nu}&=&\partial_{\mu}\hat{A}_{\nu}-\partial_{\nu}\hat{A}_{\mu}-i[\hat{A}_{\mu},\hat{A}_{\nu}]+\frac{1}{2}\theta^{\lambda\sigma}\{\partial_{\lambda}A_{\mu},\partial_{\sigma}A_{\nu}\}\nonumber\\
&&+\frac{1}{2}\theta^{\lambda\sigma}\theta_{\mu\alpha}\theta_{\nu\beta}\partial_{\lambda}\theta^{\alpha\eta}\partial_{\sigma}\theta^{\beta\delta}\{A_{\eta},A_{\delta}\}+\frac{1}{2}\theta^{\lambda\sigma}\theta_{\mu\alpha}\partial_{\lambda}\theta^{\alpha\eta}\{A_{\eta},\partial_{\sigma}A_{\nu}\}\nonumber\\
&&+\frac{1}{2}\theta^{\lambda\sigma}\theta_{\nu\beta}\partial_{\sigma}\theta^{\beta\delta}\{\partial_{\lambda}A_{\mu},A_{\delta}\}+\mathcal{O}(\theta^2)
\label{F}
\end{eqnarray}
The gauge transformation of $\hat{F}_{\mu\nu}$ is obtained from (\ref{trans}) and (\ref{45}) as,
\begin{eqnarray}
\delta_{\alpha}\hat{F}_{\mu\nu}=i[\hat{\alpha},\hat{F}_{\mu\nu}]-\frac{1}{2}\theta^{\lambda\sigma}\{\partial_{\lambda}\hat{\alpha},\partial_{\sigma}\hat{F}_{\mu\nu}\}-\frac{1}{2}\theta^{\lambda\sigma}\theta_{\mu\alpha}\theta_{\nu\beta}\partial_{\sigma}(\theta^{\alpha\eta}\theta^{\beta\delta})\{\partial_{\lambda}\hat{\alpha},\hat{F}_{\eta\delta}\}.
\end{eqnarray}
 An important consistency check is due at this point. The gauge transformation of $\hat{F}_{\mu\nu}$ can alternatively be obtained from its definition (\ref{F}) and the gauge transformation (\ref{deltaA}) of $\hat{A}_{\mu}$. For compatibility of the different expressions of $\delta_{\alpha}\hat{F}_{\mu\nu}$ we require
\begin{eqnarray}
&&-\theta^{\mu\lambda}\theta^{\nu\kappa}\partial_{\lambda}(\theta_{\kappa\sigma}\theta^{\alpha\beta}\partial_{\beta}\theta^{\sigma\delta})+\theta^{\nu\lambda}\theta^{\mu\kappa}\partial_{\lambda}(\theta_{\kappa\sigma}\theta^{\alpha\beta}\partial_{\beta}\theta^{\sigma\delta})\nonumber\\
&&+\theta^{\lambda\beta}\partial_{\lambda}\theta^{\mu\alpha}\partial_{\beta}\theta^{\nu\delta}+\theta^{\lambda\beta}\partial_{\lambda}\theta^{\mu\delta}\partial_{\beta}\theta^{\nu\alpha}=0.
\end{eqnarray}
This is satisfied if we assume the $\hat{x}$ dependent $\theta^{\mu\nu}$ according to (\ref{lie}).

 The S-W map for the field strength tensor $\hat{F}_{\mu\nu}$ can now be calculated straightforwardly by substituting the map of $\hat{A}_{\mu}$ in the defining equation (\ref{F}). It is given by
\begin{eqnarray}
\hat{F}_{\mu\nu}&=&F_{\mu\nu}+\frac{1}{2}\theta^{\lambda\sigma}\{F_{\mu\lambda},F_{\nu\sigma}\}-\frac{1}{4}\theta^{\lambda\sigma}\{A_{\lambda},(\partial_{\sigma}+\mathcal{D}_{\sigma})F_{\mu\nu}\}\nonumber\\
&&+\frac{1}{2}\theta_{\nu\lambda}\theta^{\beta\sigma}\partial_{\sigma}\theta^{\alpha\lambda}\{F_{\mu\alpha},A_{\beta}\}-\frac{1}{2}\theta_{\mu\lambda}\theta^{\beta\sigma}\partial_{\sigma}\theta^{\alpha\lambda}\{F_{\nu\alpha},A_{\beta}\}
\label{mapF}
\end{eqnarray}
where the covariant derivative is defined in the adjoint representation
\begin{eqnarray}
\mathcal{D}_{\sigma}F_{\mu\nu}=\partial_{\sigma}F_{\mu\nu}-i[A_{\sigma},F_{\mu\nu}].
\label{50}
\end{eqnarray}
Note again that in the limit of constant $\theta$ the map for $\hat{F}_{\mu\nu}$ given in (\ref{mapF}) reproduces the corresponding well known SW map\cite{SW}.
\section{Noncommutative gravity} We have now all the tools at our disposal to develop the commutative equivalent theory of noncommutative gravity in the framework of Poincare gauge gravity. As discussed in the introduction, the corresponding
 noncommutative gauge transformation can be decomposed in the following way
\begin{eqnarray}
\hat{\Lambda}(\hat{x})=\hat{\xi}^{\mu}(\hat{x})p_{\mu}+\frac{1}{2}\hat{\lambda}^{ab}(\hat{x})\Sigma_{ab}.
\label{gauge}
\end{eqnarray}
Here $\hat{\xi}^{\mu}$ is the local translation of the tetrad which must be restricted to the form given in equation (\ref{restrtrans}) in order to preserve the noncommutative algebra (\ref{lie}). The parameters $\hat{\lambda}^{ab}(\hat{x})$ characterize the local Lorentz transformations at $\hat{x}$ with $\Sigma_{ab}$
as the generators of the Lorentz group. In actual computation we have to consider some representation of these generators. In what follows we will assume the vector representation
\begin{equation}
\left[\Sigma_{cd}\right]_{ab} = \eta_{ac}\eta_{bd} - \eta_{ad}\eta_{bc}.
\label{vector}
\end{equation}
where $\eta_{ab}$ is the Minkowski metric,
\begin{eqnarray}
\eta_{ab}={\textrm{diag}}(-,+,+,+)
\end{eqnarray}
 As is usual we will denote the general coordinates by the Greek indices and components with respect to the tetrad by Latin indices.
 Corresponding to the noncommutative gauge transformations (\ref{gauge}) we introduce the gauge potential
\begin{eqnarray}
\hat{A}_{a}(\hat{x})=(\hat{D}_a)=i\hat{E}^{\mu}_a(\hat{x})p_{\mu}+\frac{i}{2}\hat{\omega}_a^{ \ bc}(\hat{x})\Sigma_{bc}
\end{eqnarray}
where $E^{\mu}_{a}(\hat{x})$ are the components of the noncommutative tetrad $\hat{E}_a$ which are also the gauge fields corresponding to general coordinate transformations and $\hat{\omega}_a^{ \ bc}(\hat{x})$ are the spin connection fields associated with local Lorentz invariance. Since $p_{\mu} = -i\partial_{\mu}$, the noncommutative tetrad maps trivially on the commutative one \cite{CK1}. Assuming the gauge transformations and the spin connection fields in the enveloping algebra we can write
\begin{eqnarray}
\hat{\Lambda}=\Lambda(x)+\Lambda^{(1)}(x,\omega_a)+\mathcal{O}(\theta^2)
\end{eqnarray}
\begin{eqnarray}
\hat{\omega}_a=\omega_a(x)+\omega_a^{(1)}(x,\omega_a)+\mathcal{O}(\theta^2)
\end{eqnarray}
where
\begin{eqnarray}
\Lambda(x)&=&\xi(x)p_{\mu}+\frac{1}{2}\lambda^{ab}(x)\Sigma_{ab}\\
\omega_a(x)&=&\frac{1}{2}\omega_a^{ \ bc}\Sigma_{bc}
\end{eqnarray}
Invoking the results (\ref{29},\ref{seibergA}) obtained in the last section we can immediately write down the order $\theta$ corrections,
\begin{eqnarray}
\Lambda^{(1)}=\frac{1}{4}\theta^{ab}\{\partial_a\Lambda,\omega_b\}
\end{eqnarray}
\begin{eqnarray}
\omega_a^{(1)}=-\frac{1}{4}\theta^{bc}\{\omega_b,\partial_c\omega_a+F_{ca}\}-\frac{1}{4}\theta_{ab}\theta^{cd}\partial_d\theta^{be}\{\omega_c,\omega_e\}
\end{eqnarray}
The field strength tensor can also be expanded in a power series of $\theta$ and we obtain from (\ref{mapF})
\begin{eqnarray}
\hat{F}_{ab}=F_{ab}+F^{(1)}_{ab}+\mathcal{O}(\theta^2)
\end{eqnarray}
where,
\begin{eqnarray}
F^{(1)}_{ab}&=&\frac{1}{2}\theta^{cd}\{F_{ac},F_{bd}\}-\frac{1}{4}\theta^{cd}\{\omega_c,(\partial_d+\mathcal{D}_d)F_{ab}\}\nonumber\\&&
+\frac{1}{2}\theta_{bc}\theta^{de}\partial_e\theta^{fc}\{F_{af},\omega_d\}-\frac{1}{2}\theta_{ac}\theta^{de}\partial_e\theta^{fc}\{F_{bf},\omega_d\}.
\label{62}
\end{eqnarray}
The field strength $F_{ab}$ in general contains both Riemann tensor $R_{ab}^{ \  \ cd}$ and the torsion $T_{ab}^{\ \ c}$. Setting the classical torsion to be zero we get
\begin{eqnarray}
F_{ab}=\frac{1}{2}R_{ab}^{ \  \ cd}\Sigma_{cd}.
\end{eqnarray}
The noncommutative Riemann Tensor $\hat{R}_{ab}^{ \  \ cd}(\hat{x})$ is obtained from 
\begin{eqnarray}
\hat{R}_{ab}(\hat{x})=\frac{1}{2}\hat{R}_{ab}^{ \  \ cd}(\hat{x})\Sigma_{cd}
\end{eqnarray}
where $\hat{R}_{ab}$ is identified with $\hat{F}_{ab}$ under the condition of zero torsion. Explicitly
\begin{eqnarray}
\hat{R}_{ab}=R_{ab}+R^{(1)}_{ab}+\mathcal{O}(\theta^2)
\end{eqnarray}
where the correction term is obtained from (\ref{62}) as,
\begin{eqnarray}
R^{(1)}_{ab}&=&\frac{1}{2}\theta^{cd}\{R_{ac},R_{bd}\}-\frac{1}{4}\theta^{cd}\{\omega_c,(\partial_d+\mathcal{D}_d)R_{ab}\}\nonumber\\&&
+\frac{1}{2}\theta_{bc}\theta^{de}\partial_e\theta^{fc}\{R_{af},\omega_d\}-\frac{1}{2}\theta_{ac}\theta^{de}\partial_e\theta^{fc}\{R_{bf},\omega_d\}.
\label{67}
\end{eqnarray}
The Ricci tensor $\hat{R}_{a}^{ \ c}=\hat{R}_{ab}^{ \  \ bc}$ and the Ricci scalar $\hat{R}=\hat{R}_{ab}^{ \  \  ab}$ are formed to construct the action
\begin{eqnarray}
S&=&\int d^4x \ \frac{1}{2\kappa^2}\hat{R}(\hat{x})\\
&=&\int d^4x \ \frac{1}{2\kappa^2}\left(R(x)+R^{(1)}(x)\right)+\mathcal{O}(\theta^2).
\end{eqnarray}
The first order correction term to the Lagrangian is
\begin{eqnarray}
R^{(1)}(x)=R^{(1)ab}_{ab}=[R^{(1)}_{ab}]^{ab}
\end{eqnarray}
It is convenient to arrange the correction as 
\begin{eqnarray}
[R^{(1)}_{ab}]^{ab}=\mathcal{R}_1+\mathcal{R}_2+\mathcal{R}_3+\mathcal{R}_4.
\label{correction}
\end{eqnarray}
where $\mathcal{R}_1,...,\mathcal{R}_4$ correspond to the contributions coming from the four pieces appearing on the r.h.s. of (\ref{67}) in the same order. It is now simple to get the first term,
\begin{eqnarray}
\mathcal{R}_1=2\theta^{cd}[R_{acg}^{ \  \  \ a}R_{bd}^{ \  \ bg}+R_{ac \ g}^{ \  \ b}R_{bd}^{ \  \ ga}].
\end{eqnarray}
For evaluating $\mathcal{R}_2$ we first compute the part containing the covariant derivative
\begin{eqnarray}
[(\partial_d+\mathcal{D}_d)R_{ab}]^e_{ \ f}=2\partial_dR_{ab \ f}^{ \  \ e}-i[\omega_d,R_{ab}]^e_{ \ f}.
\label{covder}
\end{eqnarray}
We have used the expression (\ref{50}) for the covariant derivative $\mathcal{D}_d$. Then the second correction term becomes
\begin{eqnarray}
\mathcal{R}_2&=&-\theta^{cd}\left[\frac{1}{2}(\omega_c^{ \ aj}\partial_dR_{abj}^{ \  \  \ b}-\omega_c^{ \ aj}\partial_dR_{ba \ j}^{ \  \ b})\right]\nonumber\\
&&+\frac{i}{4}\theta^{cd}\omega_c^{ \ ab}\left[\omega_{db}^{ \  \ g}R_{ajg}^{ \  \  \ j}+R_{bja}^{ \  \  \ g}\omega_{dg}^{ \  \ j}+\omega_d^{ \ jg}R_{jbga}+R_{ja}^{ \  \ jg}\omega_{dgb}\right].
\end{eqnarray}
Exploiting the various symmetries of the Riemann Tensor, spin connection and the noncommutative structure $\theta^{ab}$ we can easily show that both $\mathcal{R}_1$ and $\mathcal{R}_2$ individually vanish. Note that these terms do not depend on the coordinate dependence of $\theta^{ab}$ and will remain valid for canonical noncommutative structure. Now the last two terms on the r.h.s. of (\ref{correction}) are
\begin{eqnarray}
\mathcal{R}_3=\frac{1}{2}\theta_{jk}\theta^{nl}\partial_l\theta^{mk}[\omega_n^{ \ ab}R_{amb}^{ \  \  \  \ j}+\omega_n^{ \ aj}R_{im \ a}^{ \  \ i}]
\end{eqnarray}
and
\begin{eqnarray}
\mathcal{R}_4=-\frac{1}{2}\theta_{ik}\theta^{nl}\partial_l\theta^{mk}[\omega_n^{ \ ai}R_{jm \ a}^{ \  \ j}+\omega_n^{ \ ab}R_{amb}^{ \  \  \  \ i}]
\end{eqnarray}
Clearly these terms owe their existence to the Lie -- algebraic noncommutativity assumed in the present work. Most significantly
\begin{eqnarray}
\mathcal{R}_3+\mathcal{R}_4=0
\end{eqnarray}
as can be demonstrated easily by changing dummy variables in any one of the terms on the l.h.s. We thus find that the first order correction vanishes again for the more general structure of the noncommutative tensor assumed here.

We know that the first order correction to noncommutative gravity vanishes for constant $\theta$. Now we find that the same result holds for Lie algebraic noncommutativity. From the analysis presented here it is clear that the non existence of the order $\theta$ correction is due to various symmetries of the Riemann tensor and the spin connection of the zero order theory. It thus appears that the vanishing of the first order correction is due to the underlying symmetries of space time which will presumably hold for more general noncommutative structure. However, at this point, we dont have a definitive proof of this.
\section{Conclusions} 
A formulation of noncommutative (NC) gravity \cite{CK1} has been discussed where the coordinates satisfy a general Lie algebra. A restricted class of general coordinate transformations has been identified which preserves the noncommutative algebra. This restricted transformation is volume preserving and the corresponding theory of gravity is referred to as unimodular gravity\cite{Ein}. Our formulation of noncommutative general relativity is based on Poincare gauge gravity approach where the diffeomorphism invariance of general relativity is realized by gauging the translation of the tetrad along with localizing the Lorentz transformations with respect to the tetrad. Looking from the point of view of non -- commutative field theories the problem reduces to solving a non -- commutative Yang Mills theory where the gauge group is $ISO(3,1)$. The Seiberg -- Witten (SW) map technique \cite{SW} allows us to treat the theory as a perturbative Lagrangian theory. Since the noncommutative gauge transformations do not satisfy closure one has to take recourse of the enveloping algebra approach\cite{W1,W2,W3}. The SW maps for the noncommutative gauge parameters, potential and field strengths have been worked out in detail for the general type of noncommutativity considered here. Using these results we have computed the first order correction to NC gravity. Remarkably the first order correction is found to vanish. This shows that the vanishing of the first order correction observed for the case of canonical (constant) noncommutative algebra \cite{chams,Aschierie,MS} is more general and points to some deeper underlying connection.

As a future direction, it might be worthwhile to pursue this analysis for other formulations of NC gravity with a general noncommutative structure. This is relevant because it is known that for canonical (constant) noncommutativity nontrivial corrections begin from $\mathcal{O}(\theta^2)$ irrespective of the particular formulation of NC gravity \cite{chams,Aschierie,CK2}.


\begin{thebibliography}{99}
\bibitem{szabo1} For a recent review, see R. J. Szabo ``Symmetry, gravity and noncommutativity", {\it Class.Quant.Grav.} {\bf 23} R199 (2006) 
[hep-th/0606233]. 
\bibitem{chams} A. H. Chamseddine,  ``Deforming Einstein's gravity",
{\it Phys. Lett.} {\bf B 504} 33 (2001) [hep-th/0009153].
\bibitem{SW} N. Seiberg, E. Witten ``String theory and noncommutative geometry", {\it JHEP} {\bf 9909} 032 (1999) [hep-th/9908142].
\bibitem{Aschierie} P. Aschieri, C. Blohmann, M. Dimitrijevic, F. Meyer, P. Schupp, J. Wess ``A Gravity theory on noncommutative spaces",{\it Class.Quant.Grav.} {\bf 22} 3511-3532 (2005) [hep-th/0504183].
\bibitem{kur} S. Kurkcuoglu and C. Saemann `` Drinfeld Twist and General Relativity with Fuzzy Spaces", {\it Class. Quant. Grav.} {\bf 24} 291 (2007) [hep-th/0606197].
\bibitem{CK1} X. Calmet and A. Kobakhidze `` Noncommutative general relativity", {\it Phys. Rev. } {\bf D 72 } 045010 (2005) [hep-th/0506157].
\bibitem{Ein}J.~J.~van der Bij and H.~van Dam `` The Exchange Of Massless Spin Two Particles", {\it Physica} {\bf 116 A} 307 (1982);
F.~Wilczek `` Foundations And Working Pictures In Microphysical Cosmology", {\it Phys. Rept.} {\bf 104} 143 (1984); W.~Buchmuller and N.~Dragon `` Einstein Gravity From Restricted Coordinate Invariance", {\it Phys. Lett.} {\bf B 207} 292 (1988);
M.~Henneaux and C.~Teitelboim `` The Cosmological Constant And General Covariance", {\it Phys. Lett.} {\bf B 222} 195 (1989); W.~G.~Unruh `` A Unimodular Theory Of Canonical Quantum Gravity", {\it Phys. Rev.} {\bf D 40} 1048 (1989).
\bibitem{hari} E. Harikumar and Victor O. Rivelles `` Noncommutative Gravity", {\it Class. Quant. Grav. } {\bf 23} 7551-7560 (2006) [hep-th/0607115].
\bibitem{MS} P. Mukherjee, A. Saha `` A Note on the noncommutative correction to gravity", {\it Phys. Rev. } {\bf D 74} 027702 (2006) [hep-th/0605287].
\bibitem{Deli} C. Deliduman ``Noncommutative Gravity in Six Dimensions", 
[hep-th/0607096]. 
\bibitem{CK2} X. Calmet and A. Kobakhidze ``Second order noncommutative corrections to gravity", {\it Phys. Rev. } {\bf D 74} 047702 (2006) [hep-th/0605275].
\bibitem{Ani} 
O. Bertolami, J.G. Rosa, C.M.L. de Aragao, P. Castorina, D. Zappala ``Noncommutative gravitational quantum well", {\it
Phys. Rev.} {\bf D 72} 025010 (2005)
[hep-th/0505064];
R. Banerjee, B. Dutta Roy, S. Samanta ``Remarks on the noncommutative gravitational quantum well", {\it
Phys. Rev.} {\bf D 74} 045015 (2006) [hep-th/0605277]; A. Saha``Time-space noncommutativity in gravitational quantum well scenario", To be published in {\it Eur. Phys. J.} {\bf C} [hep-th/0609195]. 
\bibitem{W1} J. Madore, S. Schraml, P. Schupp, J. Wess `` Gauge theory on noncommutative spaces", {\it Eur. Phys. J.} {\bf C 16} 161-167 (2000) [hep-th/0001203].
\bibitem{W2} B. Jurco, S. Schraml, P. Schupp, Julius Wess `` Enveloping algebra valued gauge transformations for nonAbelian gauge groups on noncommutative spaces", {\it Eur. Phys. J. } {\bf C 17} 521-526 (2000) [hep-th/0006246].

\bibitem{W3} B. Jurco, L. Moller, S. Schraml, P. Schupp, J. Wess `` Construction of non -- Abelian gauge theories on noncommutative spaces", {\it Eur. Phys. J. } {\bf C 21} 383-388 (2001) [hep-th/0104153].
\bibitem{Behr} W. Behr, A. Sykora ``Construction of gauge theories on curved noncommutative space-time", {\it Nucl.Phys.} {\bf B 698} 473 (2004) [hep-th/0309145].

\bibitem{Weyl} H. Weyl ``Quantenmechanik und Gruppentheorie", {\it Z. Physik} {\bf 46} 1 (1927); ``The theory of groups and quantum mechanics", 
Dover, New-York (1931), translated from ``Gruppentheorie und
Quantenmechanik",
Hirzel Verlag, Leipzig (1928); E. P. Wigner ``Quantum corrections for
thermodynamic equilibrium", {\it Phys. Rev.} {\bf 40} 749 (1932). 
\bibitem{RK} R. Banerjee and K. Kumar `` Maps for currents and anomalies in noncommutative gauge theories", {\it Phys. Rev.} {\bf D 71} 045013 (2005) [hep-th/0404110].
\end{thebibliography}
\end{document}